# Nudge: Haptic Pre-Cueing to Communicate Automotive Intent

**Nikhil Gowda**
California College of the Arts
San Francisco CA 94107
ngowda@cca.edu

**Srinath Sibi**
Stanford University
Center for Design Research
Stanford CA 94305
ssibi@stanford.edu

**Sonia Baltodano**
Stanford University
Center for Design Research
Stanford CA 94305
sbaltoda@stanford.edu

**Nikolas Martelaro**
Stanford University
Center for Design Research
Stanford CA 94305
nikmart@stanford.edu

**Rohan Maheshwari**
Stanford University
Center for Design Research
Stanford CA 94305

**David Miller**
Stanford University
Center for Design Research
Stanford CA 94305
davebmiller@stanford.edu

**Wendy Ju**
Stanford University
Center for Design Research
Stanford CA 94305
wendyju@stanford.edu

## Abstract
To increase driver awareness in a fully autonomous vehicle, we developed several haptic interaction prototypes that signal what the car is planning to do next. The goal was to use haptic cues so that the driver could be situation aware but not distracted from the non-driving tasks they may be engaged in. This paper discusses the three prototypes tested and the guiding metaphor behind each concept. We also highlight the Wizard of Oz protocol adopted to test the haptic interaction prototypes and some key findings from the pilot study.

## Introduction
In an autonomous driving environment, the effectiveness of auditory and visual warning signals uan autonomous mode, the driver is likely to be distracted by tasks such as speaking on a cell phone, reading a book, or may even be asleep (supervising the car during automation causes drowsiness [1]).

It is therefore interesting to investigate whether the mechanisms underlying the detection of warnings while a driver is in control also extends to vehicle trust when it is in autonomous mode. The aim of the present study was to examine whether a haptic pre-cueing system can convey information concerning the actions that an autonomous vehicle is about to take, thereby

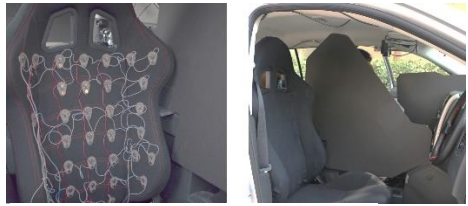
Figure 1: Vibro-tactile Back Rest

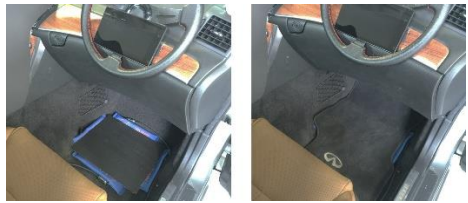
Figure 2: Dynamic Foot Displacement

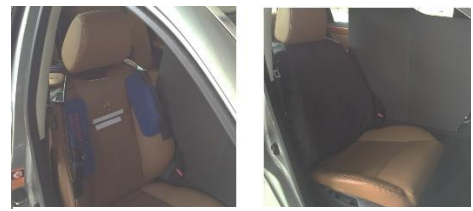
Figure 3: Dynamic Shoulder Displacement

potentially increasing the driver's awareness of the environment, while engaged in other tasks. We designed three haptic interaction prototypes and tested them in two different types of simulated autonomous cars in an on-road study (N=35).

## Previous Research/Related Work

In the past few years, there has been a great deal of interest in the development of driver assistance systems, in particular, nonvisual and multisensory collision-warning systems for drivers [2]. It has been advocated that tactile warnings are highly intuitive and may be associated with rapid responses [3].
HMI research of prototype interfaces in NHTSA Level 3 (limited self-driving automation) and NHTSA Level 4 (full self-driving automation) vehicles are currently conducted using virtual driving simulators, with necessarily some loss in fidelity. Naturalistic observation [5] of drivers in simulated environments do not yield the same results as seen in on road behavior, due to driver self-adaptation to the virtual environment. Hence, for ecological validity we ran our study on the RRADS platform [4], an autonomous vehicle simulator which uses a road vehicle with a hidden wizard driver, and the participant seated in the passenger seat, unaware of the human driver's presence.

## Design Process

We started by exploring the various ways the car could establish a haptic channel of communication to interact with the driver. We thought of metaphors that could guide our designs, such as 'riding a motorbike,' 'driving a convertible with the roof down,' or 'riding a roller coaster.' We invited colleagues to experience the interaction using improvisations and puppeteered low-fidelity prototypes. From the feedback collected we designed three prototypes.

## Prototypes

*Vibro-tactile Back Rest* (Figure 1)
Vibro-tactile stimuli was presented via 48 points of contact (2 x 10mm Shaftless Vibration Motor 3.4mm Button Type at each point). The motors are driven at an intensity that is sufficient to deliver clearly perceptible vibro-tactile stimuli. Four types of vibro-tactile stimuli are presented. The cues are in the form of bursts (15ms on, 10ms off, and 15ms on) that creates a pattern of moving intensity.

*Dynamic Foot Displacement* (Figure 2)
The second haptic device was a Pneumatic Floorboard with 4 degrees of freedom. The device consisted of 4 pneumatic air bags whose inflation was controlled by 6 gang valves to form the control manifold. An air compressor is kept in the boot of the car and its storage cylinder is used as the source of positive air pressure (40psi) to actuate the pneumatic floorboard which would tilt left-to-right or front-to-back.

*Dynamic Shoulder Displacement* (Figure 3)
The third haptic device is a two degree of freedom pneumatic shoulder actuation mechanism. The device consists of the same control manifold and air supply as the pneumatic floorboard, but only two pneumatic air bags instead of four. The air bags are attached to the seat so that they can move the participants' shoulders with air pressure.

## Pilot Test

*Participants* - We recruited 35 participants (24 female, 11 male) with a valid driving license and at least 2 years of driving experience, from mailing lists both outside and within Stanford University. The average age was 24 years. A gift card of $15 was given in return for participation.

*Methodology* - We ran the study on the RRADS platform [4] in two vehicles: A 2008 Jeep Compass and a 2012 Infiniti M45, following a traditional Wizard of Oz methodology. The RRADS involved 2 experimenters:

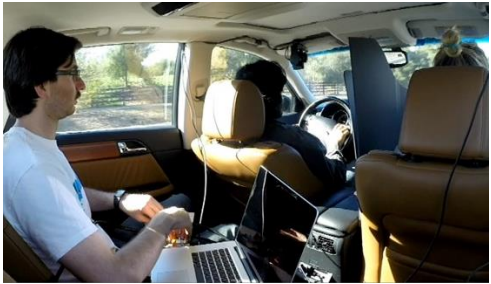
Figure 4: In car view

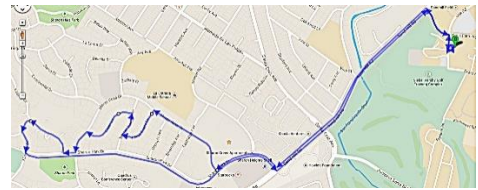
Figure 5: Route Map

one experimenter drove the vehicle while the second experimenter sat in the rear and activated the prototype devices at the appropriate times. The experimenter driving the car was hidden from view and not introduced to the participant. The second experimenter ensured that the onset of the haptic cue was as close as possible to three seconds ahead of encountering the on road event.

We employed a between-participants protocol, so each participant either interacted with one of the prototypes or were in a control condition (no feedback mechanism present). We measured the trust and likeability of the haptic devices through post-experiment questionnaires and the time from participants' response to the on road events using the quad video captures. We coded the videos of all participants at 7 events.

The seven events were specifically chosen, as just the environmental visual cues at these locations would be insufficient to determine the car's path. We coded the videos for three points of data at each event; the time of experimenter's haptic cue, the time of the participant's response (using a hand gesture) and the time of the on-road event. The time of the experimenter's haptic cue and the participant's response were then measured relative to the on-road event, to see if one prototype elicited a faster reaction time than another.

High resolution video cameras were mounted at three points within the car. The first view captured the participant's face and their responses, the second view captured the wizard's actions in the rear of the car and the third view captured the road from the just above the driver's point of view. The individual camera outputs were then fed into a quad video recorder in order to time sync the videos.

*Protocol* - At the start of the study, the experimenter controlling the interaction system was introduced to the participant. The participant then sits in the front passenger seat of the vehicle and is 'greeted' by the vehicle. This greeting is done using one of the prototype devices, or in the control case, an engine rev. The car was driven through a residential neighborhood for a period of approximately 15 minutes, averaging 25 mph. During the drive, participants encountered stop signs, traffic lights, construction vehicles, cyclists, and pedestrians. They are given a short animated movie to watch that lasts the same time as the drive. The driver (second experimenter) of the vehicle drove the same course, using the same driving pattern and speed, for all participants. At the conclusion of the course, an exit interview is conducted to collect qualitative results.

*Measures* - The key measures in this study were trust, number of correct guesses of upcoming maneuvers, timing of the gestural responses to the guessing game, and emotional reaction. Trust is measured using the questionnaire by Jian et al. to measure trust in automation technology [6]. The time from when an event was triggered and the participant made an accurate guess was measured for each event and compared between prototypes. The number of correct guesses per prototype was also calculated as a measure of clarity. We also used FACS [7] to decode facial expressions during post-analysis.

## Results

The vibration array haptic device was implemented in the Jeep Compass and the pneumatic haptic devices were tested in the Infiniti M45. The choice of cars was based on availability. We measured the trust and likeability of the haptic devices through post experiment questionnaires and coding the time from participants' response to the on road event using the quad video capture data.

The maximum difference in trust was found between the Vibration Array and Pneumatic Shoulder devices. Vibration array effectiveness is dependent on body size and geometry, and the thickness of clothing worn. Even when the stimuli was clear the vibration pattern seemed to take longer for the participant to process. It was also found that, on average, the response time of participant to the experimenter's haptic cue was fastest in the Pneumatic Floorboard and the slowest in the Vibe Array.

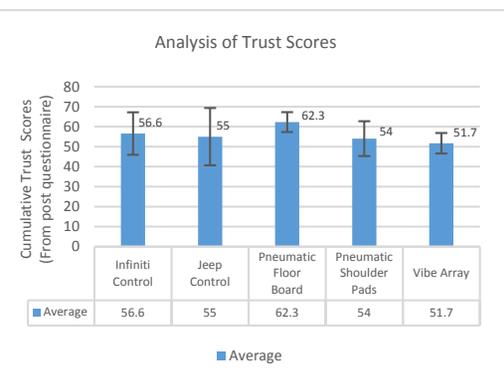

Figure 6: Analysis of trust scores from post experiment questionnaire.

It was also seen that the pneumatic shoulder pads produced the most accurate responses among all three haptic devices. However, it is also important to note that participants experiencing the pneumatic shoulder prototypes found the nature of the device to be uncomfortable as it often was sudden and caused a change in the posture of the participant and was noted to make some people angry.

We hypothesize that the small differences in trust between the designs is due to the reliable and safe driving style of the driver of the simulated autonomous vehicle. During the post interview, participants often said that they trusted the car due to its safe driving style. [Figure 6]

It is also important to note that we had initially hypothesized that in the control condition, the participant would react to an on-road event after the event occurred. This, however, was not the case. We found that environmental cues from the road and from the vehicle's motion were always present and played a significant role in the participants' guesses. Hence, the participants reacted before events occurred in many cases, similar to participants in the pre-cue conditions.

## Discussion

From the study it is clear that for a distracted driver, as in the case of automated driving, haptic cues could unobtrusively present information, engaging the driver's attention when needed. Moving forward it would be interesting to program the interaction devices to react in real time to motion of the car itself.
For researchers working on design solutions in this space access to a true automated car and its decision making algorithm would be a challenge. The RRADS platform [4] is an apt test bed for this type of research; however, real time car dynamic data collection and predictive model generation would also be needed to trigger the interaction at the right time.

## Acknowledgments
The authors would like to thank the effort and ongoing support of Shad Laws at Renault; Brian Mok, Graciela Arango, and Mishel Johns at the Center for Design Research, Stanford University.